\documentclass{svjour3}                     
\smartqed  
\usepackage{graphicx}
\newcommand{\bm}[1]{\mbox{\boldmath $#1$}}

\journalname{Few-Body Systems}

\begin{document}

\title{\boldmath Production and parity determination of $\Xi$ baryons
}

\author{\mbox{Yongseok Oh\and Kanzo~Nakayama \and~Helmut~Haberzettl}}

\institute{
    Yongseok Oh \at
    Department of Physics, Kyungpook National University,
    Daegu 702-701, Korea \\
    \email{yohphy@knu.ac.kr}
        \and
    Kanzo Nakayama \at
    Department of Physics and Astronomy, University of Georgia,
    Athens, GA 30602, USA \\
    \email{nakayama@uga.edu}
        \and
    Helmut Haberzettl \at
    Institute for Nuclear Studies and Department of Physics,
    The George Washington University, Washington, DC 20052, USA \\
    \email{helmut@gwu.edu}
}

\date{Received: date / Accepted: date}

\maketitle

\begin{abstract}
To understand the structure of hadrons and their production mechanisms, it is 
very important to be able to identify their quantum numbers. 
For investigating the spin and parity quantum numbers of $\Xi$ and
$\Omega$ baryons, in particular, it is very important to understand
strong interactions in the strangeness sector. 
At present, such quantum numbers are known for only a few of the $\Xi$ 
baryons, and even the parity of the ground state $\Xi(1318)$ has never been 
measured. 
In this article, we present a novel, model-independent way to determine the 
parity of $\Xi$ baryons solely based on symmetry considerations for the 
hadronic reaction $\bar{K} N \to K \Xi$.
\end{abstract}

\section{Introduction}

Regarding the production of strange baryons, the existing extensive studies on 
the production of $\Lambda$ and $\Sigma$ hyperons, which have strangeness
$S=-1$, have accumulated many data for various physical quantities that are 
relevant for describing the production mechanisms of $S=-1$ baryons and the 
parameters of nucleon resonances. 
By contrast, the production of multi-strangeness $\Xi$ ($S=-2$) and $\Omega$
($S=-3$) baryons has not been studied very often. 
One reason is that the processes for producing multi-strangeness
baryons from non-strangeness initial states have very small cross sections. As
a result, no significant information on multi-strangeness baryons has been
added during the past two decades~\cite{PDG12}.

Recently, however, the interest in multi-strangeness physics is increasing. 
For example, the major research programs at J-PARC, which has started its 
operation recently, include strangeness physics~\cite{Nagae07,Nagae08}. 
Since the J-PARC facility can provide energetic kaon beams, the production of 
$\Xi(1318)$ should be possible.
Furthermore, the planned upgrade of the kaon-beam energy will allow the
production of $\Xi$ resonances, thus providing a new turning point in the study 
of multi-strangeness baryons. 
Furthermore, the $\overline{\rm P}$ANDA Collaboration has a plan to investigate 
the $\bar{p} p \to \bar{\Xi} \Xi$ reaction at FAIR~\cite{SP04}, and the CLAS 
Collaboration at JLab has established the feasibility of investigating $\Xi$ 
spectroscopy via photoproduction reactions like $\gamma p \to K^+ K^+ \Xi^-$ 
and $\gamma p \to K^+ K^+ \pi^-\Xi^0$~\cite{CLAS04d,VSC12}. 
The reported data for the total and differential cross sections as well as the 
$K^+K^+$ and $K^+\Xi^-$ invariant mass distributions for the 
$\gamma p \to K^+ K^+ \Xi^-$ reaction~\cite{CLAS07b} are the first data 
measured for the exclusive photoproduction of the $\Xi$. 
Recent theoretical investigations of this reaction can be found in 
Refs.~\cite{NOH06,MON11}.

The investigation of multi-strangeness baryons will shed light on our
understanding of baryon structure by allowing us to distinguish various hadron
models. (See, for example, Ref.~\cite{Oh07} and references therein.) However,
this requires knowledge of the quantum numbers of the baryons since the spin 
and parity quantum numbers heavily depend on the internal structure of the
baryons and the underlying dynamics. 
It is of crucial importance, therefore, to identify the quantum numbers of the 
observed $\Xi$ baryons.
However, identifying the spin and parity quantum numbers of multi-strangeness 
baryons is very challenging. 
For example, the fact that the $\Omega^-(1672)$ baryon has spin-3/2 was only 
recently confirmed by experiment~\cite{BABAR06}, more than 40 years after the 
discovery of the $\Omega^-$ baryon~\cite{BCCC64}. 
For the cascade baryons, only for three states, $\Xi(1318)$, $\Xi(1530)$, and 
$\Xi(1820)$, of the eleven cascade states reported in PDG~\cite{PDG12}, the 
spin and parity quantum numbers are known. 
In a recent report~\cite{BABAR08}, the BABAR Collaboration claimed that the 
$\Xi(1690)$ has $J^P = \frac12^-$. 
What is surprising is that the parity of the cascade ground state $\Xi(1318)$ has 
never actually been measured; the positive parity quoted by the PDG~\cite{PDG12}
is taken from quark-model calculations.

In this article, we describe a novel, \textit{model-independent} way to determine 
the parity of spin-1/2 $\Xi$ baryons. 
To this end, we consider the reaction, $\bar{K}(q) + N(p) \to K(q') + \Xi(p')$,
where the arguments $q$, $p$, $q'$, and $p'$ stand for the four-momenta of the
respective particles. Although the production mechanisms are highly
model-dependent~\cite{SKL11,SST11}, our method is based on Bohr's
theorem~\cite{Bohr59} and thus a direct consequence of the rotation and
parity-inversion symmetries of the reaction amplitudes~\cite{NOH12}, and, 
therefore, model-independent. 
Furthermore, the cross section for this reaction is large enough to be measured 
experimentally.

\section{Polarization observables and the parity of spin-1/2 $\Xi$ baryons}

When the produced $\Xi$ has spin-1/2, the most general spin structure of the
reaction amplitude constrained by symmetry principles for the reaction of
$\bar{K} N \to K \Xi$ can be written as
\begin{eqnarray}
\hat{M}^+ = M_0 + M_2 \, \bm{\sigma}\cdot\hat{\bm{n}}_2^{},
\qquad 
\hat{M}^- = M_1 \, \bm{\sigma}\cdot\hat{\bm{n}}_1^{}
+ M_3 \, \bm{\sigma}\cdot\hat{\bm{q}},
\label{eq:1}
\end{eqnarray}
where $\hat{M}^+$ and $\hat{M}^-$, respectively, apply to $\Xi$ having
positive or negative parity.
The amplitude functions $M_i$ are functions of Mandelstam variables. 
The unit  vectors $\hat{\bm{n}}_1^{}$ and $\hat{\bm{n}}_2^{}$ are defined as 
$\hat{\bm{n}}_1^{} \equiv (\bm{q}\times\bm{q}')\times\bm{q} / 
|(\bm{q}\times\bm{q}')\times\bm{q}|$ and 
$\hat{\bm{n}}_2^{} \equiv \bm{q}\times\bm{q}' / |\bm{q}\times\bm{q}'|$,
respectively. 
Without loss of generality, we may choose the coordinate systems such that 
$\hat{\bm{q}}$ is the unit vector along the positive $z$-axis and
$\hat{\bm{n}}_2^{}$ along the positive $y$-axis so that $\hat{\bm{n}}_1^{}$ is
the unit vector along the positive $x$-axis. 
The reaction plane is then defined by the two vectors $\bm{q}$ and 
$\hat{\bm{n}}_1^{}$. 
This enables us to write the reaction amplitudes in a generic manner as
\begin{equation}
\hat{M} = \sum_{m=0}^3 M_m^{} \sigma_m^{}  ,
\label{eq:2}
\end{equation}
where $\sigma_0^{}$ is the $2 \times 2$ unit matrix. 
Depending on the parity of $\Xi$, either
$M_1=M_3=0$ or $M_0=M_2=0$. 
The unpolarized cross section is then given by
\begin{equation}
\frac{d\sigma}{d\Omega} \equiv
\frac{1}{2} \,\mbox{Tr}\left( \hat{M}\hat{M}^\dagger \right)
= \sum_{m=0}^3 |M_m|^2.
\label{eq:3}
\end{equation}

The most interesting spin observable related to the parity of the $\Xi$ is the
(diagonal) spin-transfer coefficient $K_{ii}$ defined by
\begin{eqnarray}
\frac{d\sigma}{d\Omega}K_{ii} \equiv \frac{1}{2} \,
\mbox{Tr}\left( \hat{M}\sigma_i^{} \hat{M}^\dagger\sigma_i^{} \right)
=  |M_0|^2 + |M_i|^2 - \sum_{k\ne i} |M_k|^2 ,
\label{eq:4}
\end{eqnarray}
for $i=1,2,3$. In terms of the cross sections, this corresponds to
\begin{equation}
K_{ii} =
\frac{\left[d\sigma_i^{}(++) + d\sigma_i^{}(--)\right]
- \left[d\sigma_i^{}(+-)+d\sigma_i^{}(-+)\right]}
{\left[d\sigma_i^{}(++) + d\sigma_i^{}(--)\right]
+ \left[d\sigma_i^{}(+-)+d\sigma_i^{}(-+)\right]}  ,
\label{eq:4a}
\end{equation}
where $d\sigma_i^{}$ stands for the differential cross section with the
polarization of the target nucleon and of the produced $\Xi$ along the
$i$-direction. The first and second $\pm$ arguments of $d\sigma_i^{}$ indicate
the parallel ($+$) or anti-parallel ($-$) spin-alignment along the
$i$-direction of the target nucleon and produced $\Xi$, respectively.

The spin-transfer coefficient $K_{ii}$ should be a function of the energy and
scattering angle, in general.
However, because of the spin structure of the amplitude, we can see that
$K_{yy}$ is constant and
\begin{equation}
K_{yy} = \pi_{\Xi}^{} .
\label{eq:5}
\end{equation}
where $\pi_{\Xi}^{}$ is the parity of the produced $\Xi$.
This is a direct consequence of the spin structures of the reaction amplitudes
as exhibited in Eq.~(\ref{eq:1}), which is a realization of reflection
symmetry in the reaction plane.
Therefore, the measurement of $K_{yy}$ will directly determine the parity of
the $\Xi$.

Another way to determine the parity of spin-1/2 $\Xi$ baryon is to use two
single-polarization observables, namely, the target-nucleon asymmetry
$T_i$ and the recoil-$\Xi$ asymmetry $P_i$, which are defined by
\begin{eqnarray}
\frac{d\sigma}{d\Omega}T_i &\equiv&
\frac{1}{2} \,\mbox{Tr}\left( M\sigma_i M^\dagger \right)
= 2\,\mbox{Re}\left[M_0 M^*_i \right]
+ 2\,\mbox{Im} \left[M_j M^*_k \right],  \\
\frac{d\sigma}{d\Omega}P_i &\equiv&
\frac{1}{2}\,\mbox{Tr}\left( M M^\dagger \sigma_i \right)
= 2\,\mbox{Re}\left[M_0M^*_i \right]
- 2\,\mbox{Im}\left[M_j M^*_k \right],
\end{eqnarray}
where the subscripts $(i,j,k)$ run cyclically.
Then the spin structures of the amplitudes of Eq.~(\ref{eq:1}) immediately
give
\begin{eqnarray}
\frac{d\sigma}{d\Omega}\left( T_y + P_y\right) =
4\,\textrm{Re}\left[ M_0 M^*_2 \right],
\qquad
\frac{d\sigma}{d\Omega}\left( T_y  - P_y\right) = 0  ,
\end{eqnarray}
for positive-parity $\Xi$ and
\begin{eqnarray}
\frac{d\sigma}{d\Omega}\left( T_y  + P_y\right) = 0  ,
\qquad
\frac{d\sigma}{d\Omega}\left( T_y - P_y\right)  =
4\,\textrm{Im}\left[ M_3M^*_1 \right],
\end{eqnarray}
for negative-parity $\Xi$.
This can be used to determine the parity of the $\Xi(1318)$ unless
$T_y \simeq P_y \simeq 0$.

\section{Summary}

In summary, we have studied the spin structure of the production amplitudes in
the reaction of $\bar{K} N \to K \Xi$. We found that the spin-transfer
coefficient $K_{yy}$, which is a double-polarization observable, can give a
direct measurement of the parity of the $\Xi$ baryon. Furthermore,  the
combinations of the two single-polarization observables, the target asymmetry
$T_y$ and the recoil asymmetry $P_y$, can give another way to check the parity
of the produced $\Xi$. Since the $\Xi$ is self-analyzing, these asymmetries can
be measured by preparing polarized nucleon targets at the current facilities
like the J-PARC. Although we considered the parity of spin-1/2 $\Xi$ baryons in
this article, our discussion can be generalized to the $\Xi$ of higher spin and
to the polarization observables in $\Xi$ photoproduction~\cite{NOH12}. Similar
relations can be found for the determination of the parity of the $\Omega$
baryons as well.

\begin{acknowledgements}
This work was partly supported by the National Research Foundation of Korea
funded by the Korean Government (Grant No.~NRF-2011-220-C00011).
The work of K.N. was also supported partly by the FFE Grant No.~41788390
(COSY-058).
\end{acknowledgements}


\end{document}